\documentclass[runningheads]{llncs}
\usepackage{graphicx}
\usepackage{multirow}
\usepackage[utf8]{inputenc} 

\begin{document}
\title{How Fake News Affect Trust in the Output of a Machine Learning System for News Curation}
\titlerunning{Trust in ML Recommendations}
%
\author{Hendrik Heuer\inst{1,2}\orcidID{0000-0003-1919-9016}\and\\
Andreas Breiter\inst{1,2}\orcidID{0000-0002-0577-8685}}
\authorrunning{Heuer \& Breiter}
%
\institute{
University of Bremen, Institute for Information Management, Germany\\
\email{\{hheuer,abreiter\}@ifib.de}
}

\maketitle              
\begin{abstract}
People are increasingly consuming news curated by machine learning (ML) systems. Motivated by studies on algorithmic bias, this paper explores which recommendations of an algorithmic news curation system users trust and how this trust is affected by untrustworthy news stories like fake news. In a study with 82 vocational school students with a background in IT, we found that users are able to provide trust ratings that distinguish trustworthy recommendations of quality news stories from untrustworthy recommendations. However, a single untrustworthy news story combined with four trustworthy news stories is rated similarly as five trustworthy news stories. The results could be a first indication that untrustworthy news stories benefit from appearing in a trustworthy context. The results also show the limitations of users' abilities to rate the recommendations of a news curation system. We discuss the implications of this for the user experience of interactive machine learning systems.

\keywords{Human-Centered Machine Learning, Algorithmic Experience, Algorithmic Bias, Fake News, Social Media}
\end{abstract}
\section{Introduction}

News curation is the complex activity of selecting and prioritizing information based on some criteria of relevance and in regards to limitations of time and space. While traditionally the domain of editorial offices of newspapers and other media outlets, this curation is increasingly performed by machine learning (ML) systems that rank the relevance of content~\cite{allcott2017social,eslami_i_2015}. This means that complex, intransparent ML systems influence the news consumption of billions of users. Pew Research Center found that around half of U.S. adults who use Facebook (53\%) think they do not understand why certain posts are included in their news feeds~\cite{pew_facebook_feed_2019}. This motivates us to explore how users perceive news recommendations and whether users can distinguish trustworthy from untrustworthy ML recommendations. We also examine whether untrustworthy news stories like fake news benefit from a trustworthy context, for instance, when an ML system predicts five stories, where four are trustworthy news stories and one is a fake news story. We operationalized the term fake news as ``fabricated information that mimics news media content in form but not in organizational process or intent''~\cite{lazer2018science}. Investigating trust and fake news in the context of an algorithmic news curation is important since such algorithms are an integral part of social media platforms like Facebook, which are a key vector of fake news distribution~\cite{allcott2017social}. Investigations of trust in news and people's propensity to believe in rumors has a long history~\cite{allport1947psychology,allport1945wartime}.

We focus on trust in a news recommender system, which connects to O'Donovan et al. and Massa and Bhattacharjee \cite{massa2004using,o2005trust}. Unlike them, our focus is not the trust in the individual items, but the trust in the ML system and its recommendations. The design of the study is shaped by how users interact with machine learning systems. Participants rate their trust in the recommendations of a machine learning system, i.e. they rate groups of news stories. Participants were told that they are interacting with an ML system, i.e. that they are not simply rating the content. We focus on trust because falling for fake news is not simply a mistake. Fake news are designed to mislead people by mimicking news media content. Our setting connects to human-in-the-loop and active machine learning, where users are interacting with a live system that they improve with their actions \cite{rubens2015active,amershi2014power,Kulesza:2015:PED:2678025.2701399}. In such settings, improving a news curation algorithm by rating individual items would require a lot of time and effort from users. We, therefore, explore explicitly rating ML recommendations as a whole as a way to gather feedback.

An investigation of how ML systems and their recommendations are perceived by users is important for those who apply algorithmic news curation and those who want to enable users to detect algorithmic bias in use. This is relevant for all human-computer interaction designers who want to enable users to interact with machine learning systems. This investigation is also relevant for ML practitioners who want to collect feedback from users on the quality of their systems or practitioners who want to crowdsource the collection of training data for their machine learning models \cite{Gulla:2017:ADN:3106426.3109436,ozgobek,Russakovsky2015}.

In our experiment, participants interacted with a simple algorithmic news curation system that presented them with news recommendations similar to a collaborative filtering system~\cite{Resnick:1994:GOA:192844.192905,herlocker2000explaining}. We conducted a between-subjects study with two phases. Our participants were recruited in a vocational school. They all had a technical background and were briefed on the type of errors that ML systems can make at unexpected times. In the first phase, participants rated their trust in different news stories. This generated a pool of news stories with trust ratings from our participants. Participants rated different subsets of news stories, i.e. each of the news stories in our investigation was rated by some users while others did not see it. In the second phase, the algorithmic news curation system combined unseen news stories for each user based on each news stories' median trust rating. This means that the trust rating of a story is based on the intersubjective agreement of the participants that rated it in the first phase. This allowed us to investigate how the trust in individual stories influences the trust in groups of news stories predicted by an ML system. We vary the number of trustworthy and untrustworthy news stories in the recommendations to study their influence on the trust rating on an 11-point rating scale. Our main goal is to understand the trust ratings of ML output as a function of the trust of individual news items for a machine learning system. In summary, this paper answers the following three research questions:

\begin{itemize}
\item Can users provide trust ratings for news recommendations of a machine learning system~(RQ1)?
\item Do users distinguish trustworthy ML recommendations from untrustworthy ML recommendations~(RQ2)?
\item Do users distinguish trustworthy ML recommendations from recommendations that include one individual untrustworthy news story~(RQ3)?
\end{itemize}

We found that users are able to give nuanced ratings of machine learning recommendations. In their trust ratings, they distinguish trustworthy from untrustworthy ML recommendations, if all stories in the output are trustworthy or if all are untrustworthy. However, participants are not able to distinguish trustworthy news recommendations from recommendations that include one fake news story. Even though they can distinguish other ML recommendations from trustworthy recommendations. 

\section{Related Work}

The goal of news recommendation and algorithmic news curation systems is to model users' interests and to recommend relevant news stories. An early example of this is GroupLens, a collaborative filtering architecture for news~\cite{Resnick:1994:GOA:192844.192905}. The prevalence of opaque and invisible algorithms that curate and recommend news motivated a variety of investigations of user awareness of algorithmic curation~\cite{Hamilton:2014:PUE:2559206.2578883,eslami_i_2015,Rader:2015:UUB:2702123.2702174,eslami2017careful}. A widely used example of such a machine learning system is Facebook's News Feed. Introduced in 2006, Facebook describes the News Feed as a ``personalized, ever-changing collection of posts from the friends, family, businesses, public figures and news sources you've connected to on Facebook''~\cite{facebook_newsfeed_2018}. By their own account, the three main signals that they use to estimate the relevance of a post are: who posted it, the type of content, and the interactions with the post. In this investigation, we primarily focus on news and fake news on social media and the impact of the machine learning system on news curation. 

Alvarado and Waern coined the term algorithmic experience as an analytic framing for making the interaction with and experience of algorithms explicit~\cite{Alvarado:2018:TAE:3173574.3173860}. Following their framework, we investigate the algorithmic experience of users of a news curation algorithm. This connects to Shou and Farkas, who investigated algorithmic news curation and the epistemological challenges of Facebook~\cite{schou2016algorithms}. They address the role of algorithms in pre-selecting what appears as representable information, which connects to our research question whether users can detect fake news stories. 

This paper extends on prior work on algorithmic bias. Eslami et al. showed that users can detect algorithmic bias during their regular usage of online hotel rating platforms and that this affects trust in the platform~\cite{eslami2017careful}. Our investigation is focused on trust as an important expression of users' beliefs. This connects to Rader et al., who explored how different ways of explaining the outputs of an algorithmic news curation system affects users' beliefs and judgments~\cite{Rader:2018:EMS:3173574.3173677}. While explanations make people more aware of how a system works, they are less effective in helping people evaluate the correctness of a system's output.


The Oxford dictionary defines trust as the firm belief in the reliability, truth, or ability of someone or something~\cite{oxford_trust}. Due to the diverse interest in trust, there are many different definitions and angles of inquiry. They range from trust as an attitude or expectation~\cite{rotter_new_1967,rempel_trust_1985}, to trust as an intention or willingness to act~\cite{mayer_integrative_1995} to trust as a result of behaviour~\cite{deutsch1960trust}. Trust was explored in a variety of different contexts, including, but not limited to intelligent systems~\cite{tullio2007works,herlocker2000explaining}, automation~\cite{muir_trust_1994,muir_trust_1996,lee2004trust}, organisations~\cite{mayer_integrative_1995}, oneself~\cite{Marsh94formalisingtrust}, and others~\cite{rotter_new_1967}. Lee and See define trust as an attitude of an agent with a goal in a situation that is characterized by some level of uncertainty and vulnerability~\cite{lee2004trust}. The sociologist Niklas Luhmann defined trust as a way to cope with risk, complexity, and a lack of system understanding~\cite{luhmann_trust_1979}. For Luhmann, trust is what allows people to face the complexity of the world. Other trust definitions cite a positive expectation of behavior and reliability~\cite{rousseau_not_1998,rotter_new_1967,muir_trust_1994}. 

Our research questions connect to Cramer et al., who investigated trust in the context of spam filters, and Berkovsky et al., who investigated trust in movie recommender systems~\cite{Berkovsky:2017:RUT:3025171.3025209,cramer_awareness_2009}. Cramer et al. found that trust guides reliance when the complexity of an automation makes a complete understanding impractical. Berkovsky et al. argue that system designers should consider grouping the recommended items using salient domain features to increase user trust, which supports earlier findings by Pu and Chen~\cite{Pu:2006:TBE:1111449.1111475}. In the context of online behavioral advertising, Eslami et al. explored how to communicate algorithmic processes by showing users why an ad is shown to them~\cite{Eslami:2018:CAP:3173574.3174006}. They found that users prefer interpretable, non-creepy explanations.

Trust ratings are central to our investigation. We use them to measure whether participants distinguish trustworthy from untrustworthy machine learning recommendations and investigate the influence of outliers. A large number of publications used trust ratings as a way to assess trust~\cite{muir_trust_1994,muir_trust_1996,MacLeod:2017:UBP:3025453.3025814,pennycook2018crowdsourcing}. In the context of online news, Pennycook and Rand showed that users can rate trust in news sources and that they can distinguish mainstream media outlets from hyperpartisan or fake news sources~\cite{pennycook2018crowdsourcing}. Muir et al. modeled trust in a machine based on interpersonal trust and showed that users can meaningfully rate their trust~\cite{muir_trust_1994}. In the context of a pasteurization plant simulation, Muir and Moray showed that operators' subjective ratings of trust provide a simple, nonintrusive insight into their use of the automation~\cite{muir_trust_1996}. Regarding the validity of such ratings, Cosley et al. showed that users of recommender system interfaces rate fairly consistently across rating scales and that they can detect systems that manipulate outputs~\cite{cosley_is_2003}. 

\section{Methods}

To explore trust in the context of algorithmic news curation, we conducted an experiment with 82 participants from a vocational school with a focus on IT. In the first phase of the study, participants with a technical background rated individual news stories, one at a time. In the second phase of the study, participants rated ML recommendations, i.e. five news stories that were presented together as the recommendations of an algorithmic news curation system. The study was conducted in situ via a web application that presented the two phases.


We recruited a homogeneous group of participants in a German vocational school. To prevent a language barrier from adding bias, the experiment was conducted in German. In Germany, the performance of students is strongly dependent on socio-economic factors \cite{pisa2006}. Students of a vocational school, which starts after compulsory schooling, have a similar background. This allows us to control for age, educational background, and socio-economic background. 
The mean age of the 82 participants was 21.40~(SD=3.92). The school had a strong STEM focus: All of the six classes were trained in IT (but they had no formal training in machine learning). The IT focus of the vocational school introduced a gender bias: 73 participants identified as male, 5 as female, 2 chose not to disclose their gender and 2 identified as a non-binary gender. This gender bias is representative of a vocational school with a STEM focus in Germany. In the training year 2016, women only accounted for 7.9\% of new IT trainees in Germany \cite{golem_fachinformatiker_2017}. 


Like Muir et al. and Cramer et al., we adopt Luhmann's definition of trust as a way to cope with risk, complexity, and a lack of system understanding~\cite{luhmann_trust_1979,muir_trust_1994,cramer_awareness_2009}. Our operationalization focuses on interpersonal and social trust, which can be described as the generalized expectancy that a person can rely on the words or promises of others~\cite{rotter_new_1967}. When consuming news, a person is making herself or himself reliant on a highly complex system that involves journalists, publishers, and interviewees. When interacting with an algorithmic news curation system, a person is making herself or himself reliant on a highly complex socio-technical system, which cannot be understood entirely and which can malfunction for myriad reasons. Each part of the system poses a risk, either due to mistakes, misunderstandings, or malicious intent. A social media platform that performs algorithmic news curation includes actors like the platform provider, the advertisers, other users, and all the different news sources with different levels of trustworthiness. All add complexity and risk. Understanding and auditing how this socio-technical system works is neither possible nor practical.


Before the experiment, we explained the rating interface, provided Mitchell's definition of ML, and briefly mentioned ML applications like object detection and self-driving cars. According to Mitchell, ``a computer program is said to learn from experience E with respect to some class of tasks T and performance measure P if its performance at tasks in T, as measured by P, improves with experience E''~\cite{Mitchell:1997:ML:541177}. To illustrate this, we showed participants how an ML algorithm learns to recognize hand-written digits. This was meant to show how and why some digits are inevitably misclassified. Algorithmic news curation was introduced as another machine learning application. The term fake news was illustrated using examples like Pope Francis backing Trump and the German Green party banning meat.

\subsection{Rating News Stories (Phase 1)}


The task in the first phase was to provide trust ratings for news stories from different sources. In this phase, participants evaluated each piece of content individually. As news stories, we used two days of publicly available Facebook posts of 13 different sources. The study was conducted in May 2017, i.e. before the Cambridge Analytica scandal and before the Russian interference in the 2016 United States elections became publicly known. 

We distinguish between seven quality media sources, e.g. public-service broadcasters and newspapers of record, and six biased sources, including tabloid media and fake news blogs. The quality media sources and the tabloid sources were selected based on their reach as measured by Facebook likes. Fake news sources were selected based on mentions in news articles on German fake news~\cite{bento_fake_2017}. Tabloid newspapers are characterized by a sensationalistic writing style and limited reliability. But, unlike fake news, they are not fabricated or intentionally misleading. For our experiment, a weighted random sample of news stories was selected from all available posts. Each of the 82 participants rated 20 news stories from a weighted random sample consisting of eight quality media news stories, four tabloid news stories, and eight fake news stories. The weighted sample accounted for the focus on fake news and online misinformation. The selected stories cover a broad range of topics, including sports like soccer, social issues like homelessness and refugees, and stories on politicians from Germany, France, and the U.S.

The presentation of the news stories resembled Facebook's official visual design. For each news story, participants saw the headline, lead paragraph, lead image, the name of the source, source logo, source URL, date and time, as well as the number of likes, comments, and shares of the Facebook post. Participants were not able to click on links or read the entire article. The data was not personalized, i.e. all participants saw the same number of likes, shares, and comments that anybody without a Facebook account would have seen if s/he would have visited the Facebook Page of the news source. In the experiment, participant rated news stories on an 11-point rating scale. The question they were asked for each news story was: ``Generally speaking, would you say that this news story can be trusted, or that you can’t be too careful? Please tell me on a score of 0 to 10, where 0 means you can’t be too careful and 10 means that this news story can be trusted''. Range and phrasing of the question are modeled after the first question of the Social Trust Scale~(STS) of the European Social Survey~(ESS) which is aimed at interpersonal trust and connected to the risk of trusting a person respectively a news story~\cite{reeskens_cross_cultural_2008}. After the experiment, the ratings of the news stories from Phase 1 were validated with media research experts. Each media researcher ranked the news sources by how trustworthy they considered the source. These rankings were compared to the median trust ratings of the news sources by the users. The experts were recruited from two German labs with a focus on media research on public communication and other cultural and social domains. All members of the two labs were contacted through internal newsletters. In a self-selection sample, nine media researcher~(three male, six female) provided their ranking via e-mail~(two from lab A, seven from lab B).

\subsection{Rating News Recommendations (Phase 2)}

\begin{figure}
\centering
  \includegraphics[width=0.55\columnwidth]{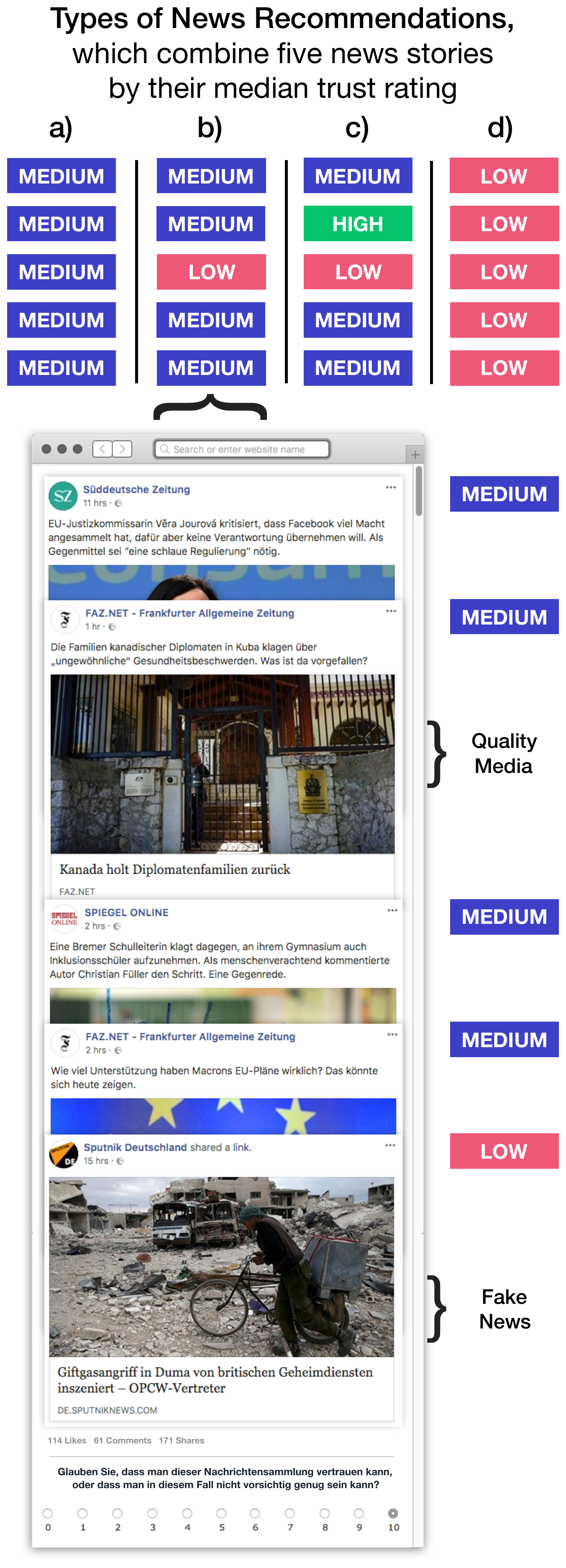}
  \caption{For Phase 2, different types of ML recommendations were generated by combining five news stories from Phase 1 by their median trust rating. Participants rated the trustworthiness of these collections of unseen news stories with a single score on an 11-point rating scale.} \label{fig:news_mixtures_types}
\end{figure}

In the second phase, participants rated their trust in the output of a news curation system. The task was not to identify individual fake news items. Participants rated the ML recommendations as a group selected by an ML system. In the study, the output of the ML system always consisted of five unseen news stories. We selected the unseen news stories based on their median trust ratings from Phase 1. The median is used as a robust measure of central tendency~\cite{huber2011robust}, which captures intersubjective agreement and which limits the influence of individual outliers. We adapted our approach from collaborative filtering systems like GroupLens~\cite{Resnick:1994:GOA:192844.192905,herlocker2000explaining}. Collaborative filtering systems identify users with similar rating patterns and use these similar users to predict unseen items. Since our sample size was limited, we couldn't train a state-of-the-art collaborative filtering system. Therefore, we used the median trust rating as a proxy.

Our goal was to understand how the presence of fake news changes the feedback users give for a machine learning system and whether trust ratings account for the presence of fake news. Our motivation was to explore how fine-grained the user feedback on a system's performance is. This is important for fields like active learning or interactive and mixed-initiative machine learning~\cite{horvitz2007reflections,amershi_power_2014,kim_2015_interactive,STUMPF2009639}, where user feedback is used to improve the system. While the experiment brief made people believe that they were interacting with a personalized ML system, the recommendations were not actually personalized. We did this to be able to compare the ratings. Unlike in Wizard of Oz experiments, there was no experimenter in the loop. Users freely interacted with an interactive software system that learned from examples.

\subsection{Types of News Recommendations}

To investigate how the trust ratings of the recommendations change based on the trustworthiness of the individual news stories, we combine five news stories in random order with different levels of trustworthiness. The scale ranges from ``can’t be too careful (0)'' to ``can be trusted (10)''. We refer to the trustworthiness of a news story as low (if the trust rating is between 0 and 3), medium~(4 to 6), and high~(7 to 10).

Figure~\ref{fig:news_mixtures_types} shows the four types of news recommendations that we discuss in this paper as well as the rating interface. 

\begin{itemize}
\item \textbf{a) Medium} --- ML output that consists of five news stories with median trust ratings between 4 and 6. 
\item \textbf{b) Medium, 1 Low} --- ML output with four news stories with ratings between 4 and 6 and one with a rating between 0 and 3 (shown in Figure~\ref{fig:news_mixtures_types}). 
\item \textbf{c) Medium, 1 Low, 1 High} --- ML output that consists of three medium news stories, one with a low trust rating and one with a high rating between 7 and 10.
\item \textbf{d) Low} --- ML output where all news stories have a trust rating between 0 and 3.
\end{itemize}

Our goal was to show as many different combinations of news recommendations to participants as possible. Unfortunately, what we were able to test depended on the news ratings in the first phase. Here, only a small subset of participants gave high ratings. This means that news recommendations like \textbf{High, 1 Low}, as well as \textbf{Low, 1 High} could not be investigated. Figure~\ref{fig:news_mixtures_types} shows the different types of ML recommendations that were presented to more than ten participants. In the figure, the five stories are shown in a collapsed view. In the experiment, participants saw each news story in its full size, i.e. the texts, images and the number of shares, likes, and comments were fully visible for each of the five news stories in the news recommendation. The news stories were presented in a web browser where participants were able to scroll. Participant rated the news recommendation on the same 11-point rating scale as the individual news items, where 0 was defined as ``you can’t be too careful'' and 10 as ``this collection of news stories can be trusted''.

\section{Results}

In Phase 1, participants were presented with individual news stories, which they rated one at a time. The news stories came from 13 different news sources. Each participant rated 20 news stories~(8 quality media, 4 tabloid, and 8 fake news stories). More than half~(53.47\%) of the trust ratings are rated as low~(with a rating between 0 and 3). 28.22\% are rated as medium~(rated 4, 5 and 6) and 18.32\% high~(7 and 10). 

The first goal of this section is to establish whether our method and the trust ratings are valid. For this, we grouped the news stories by source and ranked them by their median trust rating (Table~1). The most trustworthy news source is a conservative newspaper of record with a median trust rating of 6.0~(N=256). The least trustworthy news sources is a fake news blog with a median trust rating of 1.0~(N=129). Participants distinguish quality media~(Sources A to F) from tabloid media and fake news blogs (G to M). There is one exception: Rank H is a quality media source - produced by the public-service television - which received a median trust rating of 4.0 and which is ranked between tabloid media and fake news. Unlike the other news sources, this median trust rating is only based on one article and 25 responses. The median ratings of all other news sources are based on four or more news articles and more than 100 ratings per news source~(with a maximum of 258 ratings for 10 articles from source G). The fake news outlets are ranked as I~(9), K~(11), and M~(13).

\begin{table}
\centering
\caption{Quality media sources (marked in green) are distinguished from tabloid media (yellow) and fake news sources (red) in the Participants' Ranking ($N=82$, median trust rating) and the Media Researchers' Rankings ($N=9$).} \label{tab:sources_rating}
\includegraphics[width=.7\columnwidth]{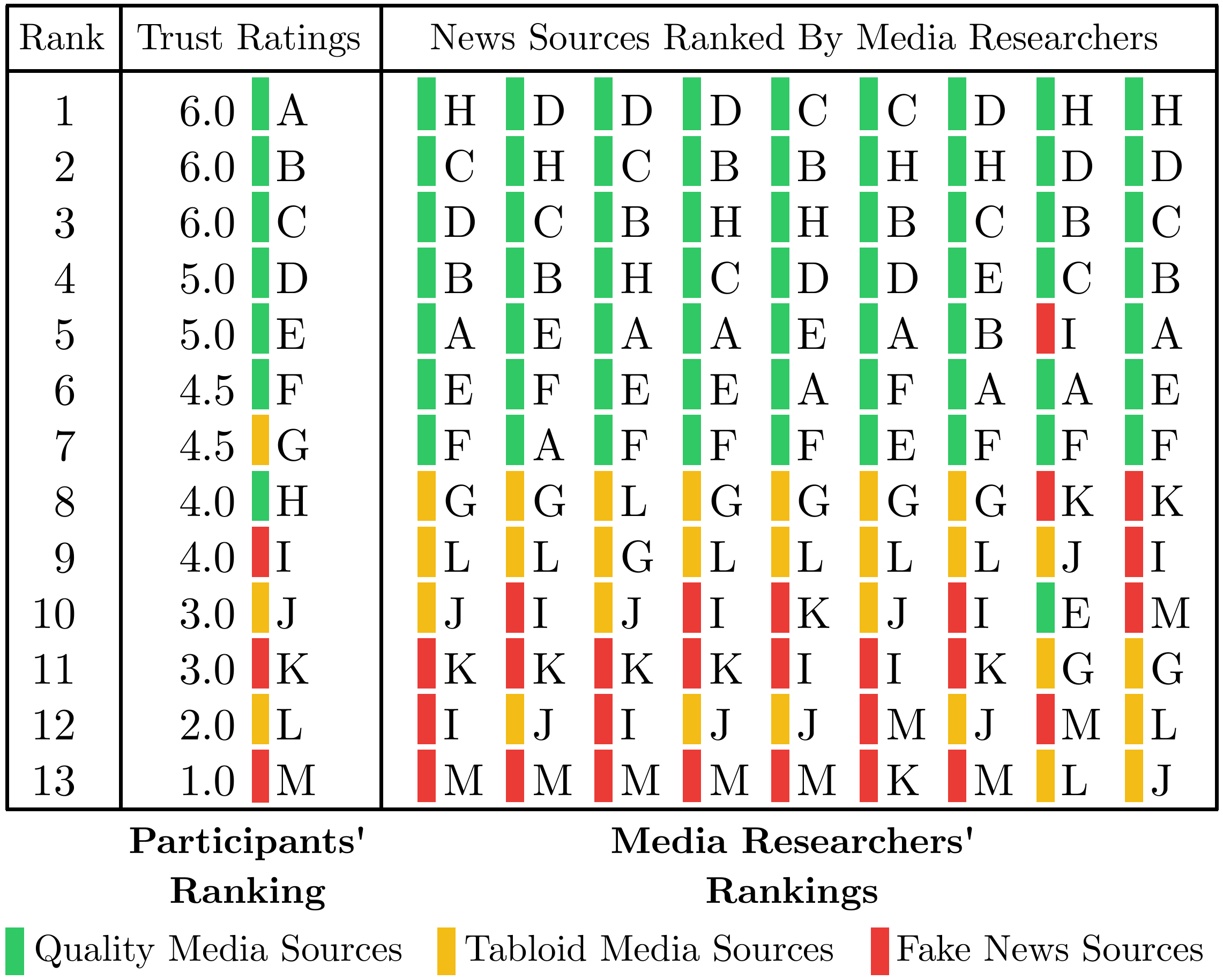}
\end{table}

We validated the trust ratings of news items by comparing them to rankings of the news sources by nine media researchers~(three male, six female), also shown in Table~1. Unlike the vocational school students, the experts did not rate individual news stories but ranked the names of the news sources by their trustworthiness. With one exception, researchers made the same distinction between quality media and biased media (fake news and tabloid media). Like our participants, the experts did not distinguish tabloid media from fake news blogs. Overall, the comparison of the two rankings shows that the trust ratings of the participants correspond to expert opinion. This validates the results through a sample different in expertise, age, and gender. The experts have a background in media research and two-thirds of the experts were female (which counterbalanced the male bias in the participants).

\subsection{Trust Ratings for Algorithmic News Curation (RQ1)}

The first research question was whether users can provide trust ratings for recommendations of an algorithmic news curation system. We addressed this question with a between-subjects design where the samples are independent, i.e. different participants saw different news stories and news recommendations~\cite{MacKenzie13}. Participants provided their trust ratings for the news stories and the news recommendations on an 11-point rating scale. We analyzed this ordinal data using a non-parametric test, i.e. we made no assumptions about the distance between the different categories. To compare the different conditions and to see whether the trust ratings of the news recommendations differ in statistically significant ways, we applied the Mann-Whitney U~test~(Wilcoxon Rank test)~\cite{mann1947test,MacKenzie13}. Like the t-test used for continuous variables, the Mann-Whitney U~test provides a p-value that indicates whether statistical differences between ordinal variables exist. Participants were told to rate ML recommendations. The framing of the experiment explicitly mentioned that they are not rating an ML system, but one recommendation of an ML system that consisted of five news stories. The results show that participants can differentiate between such ML recommendations. The ranking of the ML recommendations corresponds to the news items that make up the recommendations. Of the four types of news recommendations, a)~Medium recommendations, which consist of five news stories with a trust rating between 4 and 6, have a median rating of 5.0. d)~Low recommendations with five news stories with a low rating (0 and 3), have a median trust rating of 3.0. The trust ratings of b)~Medium, 1 Low recommendations, which combine four trustworthy stories and one untrustworthy, are rated considerably higher (4.5). ML recommendations that consist of three trustworthy news items, one untrustworthy news items (rating between 0 and 3) and one highly trustworthy news story (7 and 10), received a median trust rating of 3.0.

\subsection{Trustworthy News Recommendations (RQ2)}

\begin{figure}[t]
\centering
  \includegraphics[width=\columnwidth]{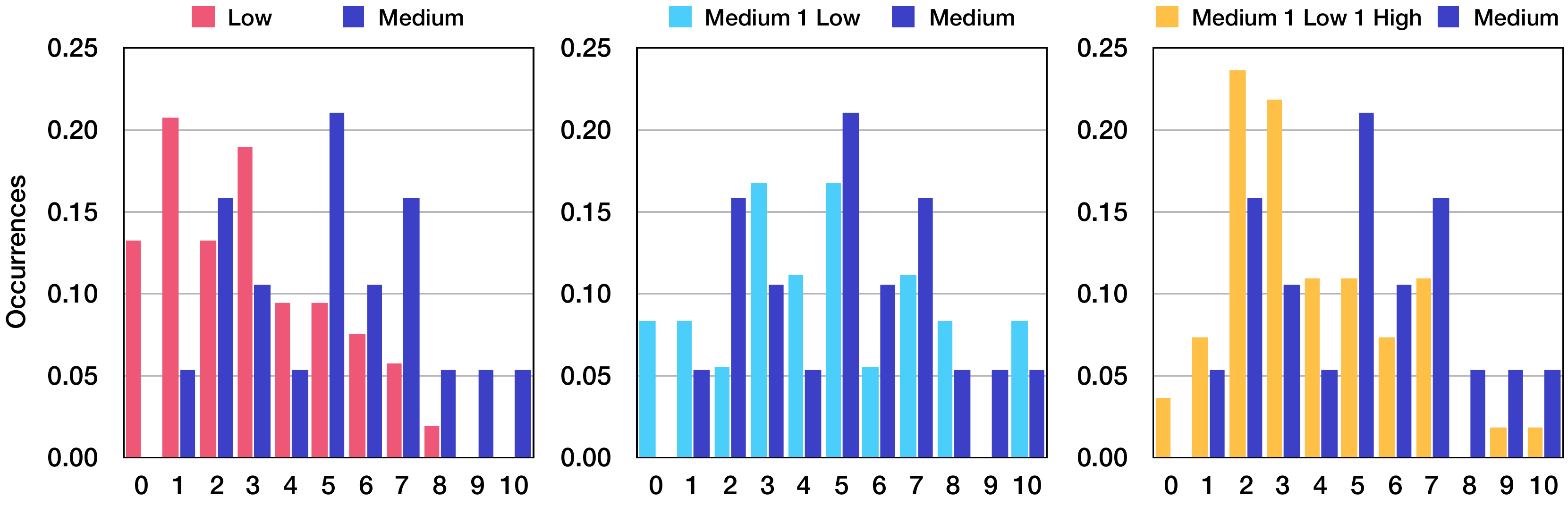}
  \caption{Histograms comparing the trust ratings of the different recommendations of the ML system.} \label{fig:histogram_ratings}
\end{figure}

\begin{table}
\centering
\caption{The Mann-Whitney U~test was applied to see whether statistically significant differences between the trust ratings of different news recommendations exist (italic for $p < 0.05$, bold for $p < 0.01$).} \label{tab:news_mixtures_test}
\begin{tabular}{|l|l|r|r|}
\hline
\multicolumn{2}{|c|}{Comparison of News Recommendations} & \multicolumn{1}{c|}{U} & \multicolumn{1}{c|}{p} \\
\hline
Medium & Low & 258.50 & \textbf{.0008} \\
Medium & M., 1 Low & 303.50 & .2491 \\
Medium & M., 1 Low, 1 High & 358.50 & \textit{.0204} \\
M., 1 Low & Low & 619.50 & \textit{.0024} \\
M., 1 Low & M., 1 Low, 1 High & 801.50 & .0618  \\
M., 1 L., 1 High & Low & 1141.50 & \textit{.0250} \\
\hline
\end{tabular}
\end{table}

The second research question was whether users can distinguish trustworthy from untrustworthy machine learning recommendations. To answer this, we compare the trust ratings of a)~Medium and d)~Low recommendations. The trustworthy a)~Medium recommendations have the same median rating (5.0) as the quality media sources D and E. Untrustworthy d)~Low recommendations with a median rating of 3.0 have the same rating as the tabloid news source J and the fake news source K. The Mann-Whitney U~test shows that participants reliably distinguish between a)~Medium and d)~Low recommendations~(U=258.5, p=.001). Figure~\ref{fig:histogram_ratings}~(left) shows the histogram of the a)~Medium recommendations, which resembles a normal distribution. 5 is the most frequent trust rating, followed by 8 and 2. The histogram of d)~Low is skewed towards negative ratings. Here, 1 and 3 are the most frequent trust rating. Nevertheless, a large number of participants still gave a rating of 6 or higher for d)~Low recommendations. A large fraction also gave a)~Medium recommendations a rating lower than 5. 

\subsection{Fake News Stories (RQ3)}

The first two research questions showed that technically advanced participants are able to differentiate between trustworthy and untrustworthy ML recommendations in an experiment where they are primed to pay attention to individual fake news stories. The most important research question, however, was whether users distinguish trustworthy ML recommendations from recommendations that include one fake news story in their ratings. For this, we compare the trust ratings of a)~Medium recommendations to those of b)~4 Medium, 1 Low recommendations, which have a median trust rating of 4.5~(N=36). Compared to a)~Medium at 5.0~(N=19), the median is slightly lower. Compared to the news sources, b)~4 Medium, 1 Low at 4.5 is similar to quality media~(Source F) and tabloid media~(Source G). The Mann-Whitney U~test shows that the ratings for b)~Medium, 1 Low recommendations are significantly different from d)~Low recommendations (U=619.5, p=.002). However, the difference between a)~Medium and b)~4 Medium, 1 Low is not statistically significant~(U=303.5, p=.249). This means that the crucial fake news case, where a recommendation consists of four trustworthy news stories and one fake news story, is not distinguished in a statistically significant way. The histogram in Figure~\ref{fig:histogram_ratings}~(center) shows that a)~Medium and b)~Medium, 1 Low are very similar. Both resemble a normal distribution and both have strong peaks at 5, the neutral position of the 11-point rating scale. a)~Medium recommendations have strong peaks at 2 and 7, b) Medium, 1 Low recommendations have peaks at 3 and 7. To see whether participants are able to distinguish the fake news case from other recommendations, we also compare b)~4 Medium, 1 Low recommendations to c)~Medium, 1 Low, 1 High recommendations, which consist of three trustworthy news stories (rated between 4 and 6), one highly trustworthy story~(7 and 10) and one untrustworthy news item~(0 and 3). The c)~3 Medium, 1 Low, 1 High recommendations are rated as 3.0~(N=55). This is the same as d)~Low recommendations~(3.0). It is also much lower than the ratings of b)~Medium, 1 Low recommendations~(4.5). In comparison to the median trust rating of the news sources, this places c)~3 Medium, 1 Low, 1 High between the tabloid source J and the fake news source K. According to the Mann-Whitney U~test, participants are able to distinguish c)~3 Medium, 1 Low, 1 High recommendations from a)~Medium~(U=358.5, p=.020) and d)~Low~(U=1141.5, p=.025) recommendations. c)~3 Medium, 1 Low, 1 High recommendations are not distinguished from the fake news case of c)~Medium, 1 Low recommendations~(U=801.50, p=.062). Figure~\ref{fig:histogram_ratings}~(right) compares the histograms of a)~Medium and c)~3 Medium, 1 Low, 1 High recommendations. The largest peaks for c)~recommendations are at 2 and 3, with very few high ratings of 7, 8, 9 or 10, but also few ratings of 0 and 1. The difference between the ratings of the two recommendations is clearly recognizable in the histograms.

\section{Discussion}

The study found that participants with a technical background can provide plausible trust ratings for individual news items as well as for groups of news items presented as the recommendations of an ML system. The ratings of the news recommendations correspond to the news stories that are part of the news recommendations. We further showed that the trust ratings for individual news items correspond to expert opinion. Vocational school students and media researchers both distinguish news stories of quality media sources from biased sources. Neither experts nor participants placed the fake news sources at the end of the rankings. These findings are highly problematic considering the nature of fake news. Following Lazer et al.'s definition of fake news as fabricated information that mimics news media content in form but not in organizational process or intent~\cite{lazer2018science}, fake news are more likely to emulate tabloid media in form and content than quality media.

We found that users can provide trust ratings for an algorithmic news curation system when presented with recommendations of a machine learning system. Participants were able to assign trust ratings that differentiated between news recommendations in a statistically significant way, at least when comparing trustworthy from untrustworthy machine learning recommendations. However, the crucial fake news case was not distinguished from trustworthy recommendations. This is noteworthy since the first phase of our study showed that users are able to identify individual fake news stories. When providing trust ratings for groups of news items in the second phase, the presence of fake news did not affect the trust ratings of the output as a whole. This is surprising since prior research on trust in automation reliance implies that user's assessment of a system changes when the system makes mistakes~\cite{dzindolet2003role}. Dzindolet et al. report that the consequences of this were so severe that after encountering a system that makes mistakes, participants distrusted even reliable aids. In our study, one fake news story did not affect the trust rating in such a drastic way. An untrustworthy fake news story did not lead to a very low trust rating for the news recommendation as a whole. The simplest explanation for this would be that the task is too hard for users. Identifying a lowly trusted news story in the recommendations of an algorithmic news curation system may overstrain users. A contrary indication against this explanation is that trustworthy and untrustworthy recommendations can be distinguished from other news recommendations like the c)~Medium, 1 Low, 1 High recommendations. 

Our findings could, therefore, be a first indication that untrustworthy news stories benefit from appearing in a trustworthy context. Our findings are especially surprising considering that the users have an IT background and were primed to be suspicious. If users implicitly trust fake news that appear in a trustworthy context, this would have far-reaching consequences. Especially since social media is becoming the primary news sources for a large group of people~\cite{newman2017reuters}. The question whether untrustworthy news stories like fake news benefit from a trustworthy context is directly connected to research on algorithmic experience and the user awareness of algorithmic curation. 

Our understanding of the user experience of machine learning systems is only emerging~\cite{Hamilton:2014:PUE:2559206.2578883,eslami_i_2015,Rader:2015:UUB:2702123.2702174,eslami2017careful}. In the context of an online hotel rating platforms, Eslami et al. found that users can detect algorithmic bias during their regular usage of a service and that this bias affects trust in the platform~\cite{eslami2017careful}. The question, therefore, is why participants did not react to the fake news stories in our study in a similar way. Further research has to show what role the context of our study - machine learning and algorithmic news curation - may have played. While framing effects are known to affect trust, our expectation was that the framing would have primed users to be overly cautious~\cite{MacLeod:2017:UBP:3025453.3025814}. This would mean that participants can distinguish them in the experiment, but not in the practice. This was not the case.

In the instructions of the controlled experiment, we define the terms fake news and machine learning. This increased algorithmic awareness and the expectation of algorithmic bias. It could also have influenced the perception and actions of the participants by making them more cautious and distrusting. We show that despite this priming and framing, participants were not able to provide ratings that reflect the presence of fake news stories in the output. If people with a technical background and a task framed like this are unable to do this, how could a layperson? Especially considering that participants were able to distinguish uniformly trustworthy from uniformly untrustworthy output. All this makes the implications of our experiment on the UX of machine learning and how feedback/training data needs to be collected especially surprising and urgent. This adds to a large body of research on algorithmic experience and algorithmic awareness~\cite{Eslami:2018:CAP:3173574.3174006,Woodruff:2018:QEP:3173574.3174230,Binns:2018:RHP:3173574.3173951}. 

\section{Limitations}

Studying trust in machine learning systems for news curation is challenging. We had to simplify a complex socio-technical system. Our approach connects to a large body of research that applies trust ratings to study complex phenomena~\cite{muir_trust_1994,muir_trust_1996,MacLeod:2017:UBP:3025453.3025814,pennycook2018crowdsourcing}. Since no ground truth data on the trustworthiness of different news stories was available, we designed a study that used the median trust ratings of our participants as intersubjective agreement on the perceived trustworthiness of a news story. A real-world algorithmic news curation system is more complex and judges the relevance of postings based on three factors: who posted it, the type of content, and the interactions with the post~\cite{facebook_newsfeed_2018}. Even though we recreated the design of Facebook's News Feed, our setting was artificial. Interactions with the posts were limited, participants did not select the news sources themselves and they did not see the likes, shares, and comments of their real Facebook ``friends''. We focused on news stories and did not personalize the recommendations of the ML system. Further research could investigate how the different sources affect the trust perception of news stories respectively the trust perception of ML recommendations. However, not personalizing the results and focusing on news was necessary to get comparable results.

We conducted the experiment in a German vocational school with an IT focus. This limits biasing factors like age, educational background, and socio-economic background, but led to a strong male bias. We counteracted this bias by validating the trust ratings of news stories with nine media research experts - a heterogeneous group that is different in age, gender~(three male, six female), and background, which confirmed our results. Prior research also implies that the findings from our sample of participants are generalizable despite the strong male bias. A German study~(N=1,011) from 2017 showed that age and gender have little influence on experience with fake news, which is similar for all people under 60, especially between 14-to-24-year olds and 25-to-44-year olds~\cite{lfm_fake_2017}. The participants in this study had a background in IT, which could have influenced the results. Prior work on algorithmically generated image captions showed that technical proficiency and education level do not influence trust ratings~\cite{MacLeod:2017:UBP:3025453.3025814}. Moreover, even if the technical background of the participants would have helped the task, they were not able to provide nuanced ratings that accounted for untrustworthy news items, which further supports our arguments.
 
\section{Conclusion}

Our study investigated how fake news affect trust in the output of a machine learning system for news curation. Our results show that participants distinguish trustworthy from untrustworthy ML recommendations in significantly different trust ratings. Meanwhile, the crucial fake news case, where an individual fake news story appears among trustworthy news stories, is not distinguished from trustworthy ML recommendations. Since ML systems make a variety of errors that can be subtle, it is important to incorporate user feedback on the performance of the system. Our study shows that gathering such feedback is challenging. While participants are able to distinguish exclusively trustworthy from untrustworthy recommendations, they do not account for subtle but crucial differences like fake news. Our recommendations for those who want to apply machine learning is, therefore, to evaluate how well users can give feedback before training active learning and human-in-the-loop machine learning systems. Further work in other real-world scenarios is needed, especially since news recommendation systems are constantly changing.

%
%
%
%
\bibliographystyle{splncs04}
\bibliography{references}

\begin{thebibliography}{10}
\providecommand{\url}[1]{\texttt{#1}}
\providecommand{\urlprefix}{URL }
\providecommand{\doi}[1]{https://doi.org/#1}

\bibitem{golem_fachinformatiker_2017}
Fachinformatiker: {IT}-{Berufsausbildung} auf dem {Arbeitsmarkt} sehr gefragt -
  {Golem}.de (2017),
  \url{https://www.golem.de/news/fachinformatiker-it-berufsausbildung-auf-dem-arbeitsmarkt-sehr-gefragt-1702-126214.html}

\bibitem{pew_facebook_feed_2019}
Many {Facebook} users don’t understand its news feed (2019),
  \url{http://www.pewresearch.org/fact-tank/2018/09/05/many-facebook-users-dont-understand-how-the-sites-news-feed-works/}

\bibitem{allcott2017social}
Allcott, H., Gentzkow, M.: Social media and fake news in the 2016 election.
  Tech. rep., National Bureau of Economic Research (2017)

\bibitem{allport1945wartime}
Allport, F.H., Lepkin, M.: Wartime rumors of waste and special privilege: why
  some people believe them. The Journal of Abnormal and Social Psychology
  \textbf{40}(1), ~3 (1945)

\bibitem{allport1947psychology}
Allport, G.W., Postman, L.: The psychology of rumor.  (1947)

\bibitem{Alvarado:2018:TAE:3173574.3173860}
Alvarado, O., Waern, A.: Towards algorithmic experience: Initial efforts for
  social media contexts. In: Proceedings of the 2018 CHI Conference on Human
  Factors in Computing Systems. pp. 286:1--286:12. CHI '18, ACM, New York, NY,
  USA (2018). \doi{10.1145/3173574.3173860},
  \url{http://doi.acm.org/10.1145/3173574.3173860}

\bibitem{amershi2014power}
Amershi, S., Cakmak, M., Knox, W.B., Kulesza, T.: Power to the people: The role
  of humans in interactive machine learning. AI Magazine  \textbf{35}(4),
  105--120 (2014)

\bibitem{amershi_power_2014}
Amershi, S., Cakmak, M., Knox, W.B., Kulesza, T.: Power to the people: {The}
  role of humans in interactive machine learning. AI Magazine  \textbf{35}(4),
  105--120 (2014)

\bibitem{bento_fake_2017}
bento, Katharina~Hölter, S.L.: Fake {News} in {Deutschland}: {Diese}
  {Webseiten} machen {Stimmung} gegen {Merkel} (2017),
  \url{http://www.bento.de/today/fake-news-in-deutschland-diese-seiten-machen-stimmung-gegen-merkel-1126168/}

\bibitem{Berkovsky:2017:RUT:3025171.3025209}
Berkovsky, S., Taib, R., Conway, D.: How to recommend?: User trust factors in
  movie recommender systems. In: Proceedings of the 22Nd International
  Conference on Intelligent User Interfaces. pp. 287--300. IUI '17, ACM, New
  York, NY, USA (2017). \doi{10.1145/3025171.3025209},
  \url{http://doi.acm.org/10.1145/3025171.3025209}

\bibitem{Binns:2018:RHP:3173574.3173951}
Binns, R., Van~Kleek, M., Veale, M., Lyngs, U., Zhao, J., Shadbolt, N.: 'it's
  reducing a human being to a percentage': Perceptions of justice in
  algorithmic decisions. In: Proceedings of the 2018 CHI Conference on Human
  Factors in Computing Systems. pp. 377:1--377:14. CHI '18, ACM, New York, NY,
  USA (2018). \doi{10.1145/3173574.3173951},
  \url{http://doi.acm.org/10.1145/3173574.3173951}

\bibitem{cosley_is_2003}
Cosley, D., Lam, S.K., Albert, I., Konstan, J.A., Riedl, J.: Is {Seeing}
  {Believing}?: {How} {Recommender} {System} {Interfaces} {Affect} {Users}'
  {Opinions}. In: Proceedings of the {SIGCHI} {Conference} on {Human} {Factors}
  in {Computing} {Systems}. pp. 585--592. {CHI} '03, ACM, New York, NY, USA
  (2003). \doi{10.1145/642611.642713},
  \url{http://doi.acm.org/10.1145/642611.642713}

\bibitem{cramer_awareness_2009}
Cramer, H.S., Evers, V., van Someren, M.W., Wielinga, B.J.: Awareness,
  {Training} and {Trust} in {Interaction} with {Adaptive} {Spam} {Filters}. In:
  Proceedings of the {SIGCHI} {Conference} on {Human} {Factors} in {Computing}
  {Systems}. pp. 909--912. {CHI} '09, ACM, New York, NY, USA (2009).
  \doi{10.1145/1518701.1518839},
  \url{http://doi.acm.org/10.1145/1518701.1518839}

\bibitem{deutsch1960trust}
Deutsch, M.: Trust, trustworthiness, and the f scale. The Journal of Abnormal
  and Social Psychology  \textbf{61}(1), ~138 (1960)

\bibitem{oxford_trust}
{Dictionaries}, O.: trust (2018),
  \url{https://en.oxforddictionaries.com/definition/trust}

\bibitem{dzindolet2003role}
Dzindolet, M.T., Peterson, S.A., Pomranky, R.A., Pierce, L.G., Beck, H.P.: The
  role of trust in automation reliance. International journal of human-computer
  studies  \textbf{58}(6),  697--718 (2003)

\bibitem{Eslami:2018:CAP:3173574.3174006}
Eslami, M., Krishna~Kumaran, S.R., Sandvig, C., Karahalios, K.: Communicating
  algorithmic process in online behavioral advertising. In: Proceedings of the
  2018 CHI Conference on Human Factors in Computing Systems. pp. 432:1--432:13.
  CHI '18, ACM, New York, NY, USA (2018). \doi{10.1145/3173574.3174006},
  \url{http://doi.acm.org/10.1145/3173574.3174006}

\bibitem{eslami_i_2015}
Eslami, M., Rickman, A., Vaccaro, K., Aleyasen, A., Vuong, A., Karahalios, K.,
  Hamilton, K., Sandvig, C.: "i always assumed that i wasn't really that close
  to [her]": Reasoning about invisible algorithms in news feeds. In:
  Proceedings of the 33rd Annual ACM Conference on Human Factors in Computing
  Systems. pp. 153--162. CHI '15, ACM, New York, NY, USA (2015).
  \doi{10.1145/2702123.2702556},
  \url{http://doi.acm.org/10.1145/2702123.2702556}

\bibitem{eslami2017careful}
Eslami, M., Vaccaro, K., Karahalios, K., Hamilton, K.: "be careful; things can
  be worse than they appear": Understanding biased algorithms and users'
  behavior around them in rating platforms. In: ICWSM. pp. 62--71 (2017)

\bibitem{facebook_newsfeed_2018}
Facebook: Facebook {News} {Feed} (2018), \url{https://newsfeed.fb.com/}

\bibitem{Gulla:2017:ADN:3106426.3109436}
Gulla, J.A., Zhang, L., Liu, P., \"{O}zg\"{o}bek, O., Su, X.: The adressa
  dataset for news recommendation. In: Proceedings of the International
  Conference on Web Intelligence. pp. 1042--1048. WI '17, ACM, New York, NY,
  USA (2017). \doi{10.1145/3106426.3109436},
  \url{http://doi.acm.org/10.1145/3106426.3109436}

\bibitem{Hamilton:2014:PUE:2559206.2578883}
Hamilton, K., Karahalios, K., Sandvig, C., Eslami, M.: A path to understanding
  the effects of algorithm awareness. In: CHI '14 Extended Abstracts on Human
  Factors in Computing Systems. pp. 631--642. CHI EA '14, ACM, New York, NY,
  USA (2014). \doi{10.1145/2559206.2578883},
  \url{http://doi.acm.org/10.1145/2559206.2578883}

\bibitem{herlocker2000explaining}
Herlocker, J.L., Konstan, J.A., Riedl, J.: Explaining collaborative filtering
  recommendations. In: Proceedings of the 2000 ACM conference on Computer
  supported cooperative work. pp. 241--250. ACM (2000)

\bibitem{horvitz2007reflections}
Horvitz, E.J.: Reflections on challenges and promises of mixed-initiative
  interaction. AI Magazine  \textbf{28}(2), ~3 (2007)

\bibitem{huber2011robust}
Huber, P.J.: Robust statistics. In: International Encyclopedia of Statistical
  Science, pp. 1248--1251. Springer (2011)

\bibitem{kim_2015_interactive}
Kim, B.: Interactive and interpretable machine learning models for human
  machine collaboration. Ph.D. thesis, Massachusetts Institute of Technology
  (2015)

\bibitem{Kulesza:2015:PED:2678025.2701399}
Kulesza, T., Burnett, M., Wong, W.K., Stumpf, S.: Principles of explanatory
  debugging to personalize interactive machine learning. In: Proceedings of the
  20th International Conference on Intelligent User Interfaces. pp. 126--137.
  IUI '15, ACM, New York, NY, USA (2015). \doi{10.1145/2678025.2701399},
  \url{http://doi.acm.org/10.1145/2678025.2701399}

\bibitem{lfm_fake_2017}
{Landesanstalt für Medien NRW (LfM)}: Fake news. Tech. rep., forsa (May 2017),
  \url{https://bit.ly/2ya2gj0}

\bibitem{lazer2018science}
Lazer, D.M., Baum, M.A., Benkler, Y., Berinsky, A.J., Greenhill, K.M., Menczer,
  F., Metzger, M.J., Nyhan, B., Pennycook, G., Rothschild, D., et~al.: The
  science of fake news. Science  \textbf{359}(6380),  1094--1096 (2018)

\bibitem{lee2004trust}
Lee, J.D., See, K.A.: Trust in automation: Designing for appropriate reliance.
  Human Factors: The Journal of the Human Factors and Ergonomics Society
  \textbf{46}(1),  50--80 (2004)

\bibitem{luhmann_trust_1979}
Luhmann, N.: Trust and power. 1979. John Willey \& Sons  (1979)

\bibitem{MacKenzie13}
MacKenzie, I.S.: Human-Computer Interaction: An Empirical Research Perspective.
  Morgan Kaufmann, Amsterdam (2013),
  \url{http://www.sciencedirect.com/science/book/9780124058651}

\bibitem{MacLeod:2017:UBP:3025453.3025814}
MacLeod, H., Bennett, C.L., Morris, M.R., Cutrell, E.: Understanding blind
  people's experiences with computer-generated captions of social media images.
  In: Proceedings of the 2017 CHI Conference on Human Factors in Computing
  Systems. pp. 5988--5999. CHI '17, ACM, New York, NY, USA (2017).
  \doi{10.1145/3025453.3025814},
  \url{http://doi.acm.org/10.1145/3025453.3025814}

\bibitem{mann1947test}
Mann, H.B., Whitney, D.R.: On a test of whether one of two random variables is
  stochastically larger than the other. The annals of mathematical statistics
  pp. 50--60 (1947)

\bibitem{Marsh94formalisingtrust}
Marsh, S.P.: Formalising trust as a computational concept. Ph.D. thesis (1994)

\bibitem{massa2004using}
Massa, P., Bhattacharjee, B.: Using trust in recommender systems: an
  experimental analysis. In: International Conference on Trust Management. pp.
  221--235. Springer (2004)

\bibitem{mayer_integrative_1995}
Mayer, R.C., Davis, J.H., Schoorman, F.D.: An {Integrative} {Model} of
  {Organizational} {Trust}. The Academy of Management Review  \textbf{20}(3),
  709--734 (1995). \doi{10.2307/258792},
  \url{http://www.jstor.org/stable/258792}

\bibitem{Mitchell:1997:ML:541177}
Mitchell, T.M.: Machine Learning. McGraw-Hill, Inc., New York, NY, USA, 1 edn.
  (1997)

\bibitem{muir_trust_1996}
Muir, B.M., Moray, N.: Trust in automation. {Part} {II}. {Experimental} studies
  of trust and human intervention in a process control simulation. Ergonomics
  \textbf{39}(3),  429--460 (Mar 1996). \doi{10.1080/00140139608964474}

\bibitem{muir_trust_1994}
Muir, B.M.: Trust in automation: {Part} {I}. {Theoretical} issues in the study
  of trust and human intervention in automated systems. Ergonomics
  \textbf{37}(11),  1905--1922 (Nov 1994). \doi{10.1080/00140139408964957},
  \url{http://dx.doi.org/10.1080/00140139408964957}

\bibitem{newman2017reuters}
Newman, N., Fletcher, R., Kalogeropoulos, A., Levy, D.A., Nielsen, R.K.:
  Reuters institute digital news report 2017  (2017),
  \url{https://ssrn.com/abstract=3026082}

\bibitem{o2005trust}
O'Donovan, J., Smyth, B.: Trust in recommender systems. In: Proceedings of the
  10th international conference on Intelligent user interfaces. pp. 167--174.
  ACM (2005)

\bibitem{pisa2006}
OECD: PISA 2006 (2007),
  \url{{https://www.oecd-ilibrary.org/content/publication/9789264040014-en}}

\bibitem{pennycook2018crowdsourcing}
Pennycook, G., Rand, D.G.: Crowdsourcing judgments of news source quality
  (2018)

\bibitem{Pu:2006:TBE:1111449.1111475}
Pu, P., Chen, L.: Trust building with explanation interfaces. In: Proceedings
  of the 11th International Conference on Intelligent User Interfaces. pp.
  93--100. IUI '06, ACM, New York, NY, USA (2006).
  \doi{10.1145/1111449.1111475},
  \url{http://doi.acm.org/10.1145/1111449.1111475}

\bibitem{Rader:2018:EMS:3173574.3173677}
Rader, E., Cotter, K., Cho, J.: Explanations as mechanisms for supporting
  algorithmic transparency. In: Proceedings of the 2018 CHI Conference on Human
  Factors in Computing Systems. pp. 103:1--103:13. CHI '18, ACM, New York, NY,
  USA (2018). \doi{10.1145/3173574.3173677},
  \url{http://doi.acm.org/10.1145/3173574.3173677}

\bibitem{Rader:2015:UUB:2702123.2702174}
Rader, E., Gray, R.: Understanding user beliefs about algorithmic curation in
  the facebook news feed. In: Proceedings of the 33rd Annual ACM Conference on
  Human Factors in Computing Systems. pp. 173--182. CHI '15, ACM, New York, NY,
  USA (2015). \doi{10.1145/2702123.2702174},
  \url{http://doi.acm.org/10.1145/2702123.2702174}

\bibitem{reeskens_cross_cultural_2008}
Reeskens, T., Hooghe, M.: Cross-cultural measurement equivalence of generalized
  trust. {Evidence} from the {European} {Social} {Survey} (2002 and 2004).
  Social Indicators Research  \textbf{85}(3),  515--532 (Feb 2008).
  \doi{10.1007/s11205-007-9100-z},
  \url{http://link.springer.com/article/10.1007/s11205-007-9100-z}

\bibitem{rempel_trust_1985}
Rempel, J.K., Holmes, J.G., Zanna, M.P.: Trust in close relationships. Journal
  of personality and social psychology  \textbf{49}(1), ~95 (1985)

\bibitem{Resnick:1994:GOA:192844.192905}
Resnick, P., Iacovou, N., Suchak, M., Bergstrom, P., Riedl, J.: Grouplens: An
  open architecture for collaborative filtering of netnews. In: Proceedings of
  the 1994 ACM Conference on Computer Supported Cooperative Work. pp. 175--186.
  CSCW '94, ACM, New York, NY, USA (1994). \doi{10.1145/192844.192905},
  \url{http://doi.acm.org/10.1145/192844.192905}

\bibitem{rotter_new_1967}
Rotter, J.B.: A new scale for the measurement of interpersonal trust. Journal
  of Personality  \textbf{35}(4),  651--665 (Dec 1967).
  \doi{10.1111/j.1467-6494.1967.tb01454.x},
  \url{http://onlinelibrary.wiley.com/doi/10.1111/j.1467-6494.1967.tb01454.x/abstract}

\bibitem{rousseau_not_1998}
Rousseau, D.M., Sitkin, S.B., Burt, R.S., Camerer, C.: Not {So} {Different}
  {After} {All}: {A} {Cross}-{Discipline} {View} {Of} {Trust}. Academy of
  Management Review  \textbf{23}(3),  393--404 (Jul 1998).
  \doi{10.5465/AMR.1998.926617}, \url{http://amr.aom.org/content/23/3/393}

\bibitem{rubens2015active}
Rubens, N., Elahi, M., Sugiyama, M., Kaplan, D.: Active learning in recommender
  systems. In: Recommender systems handbook, pp. 809--846. Springer (2015)

\bibitem{Russakovsky2015}
Russakovsky, O., Deng, J., Su, H., Krause, J., Satheesh, S., Ma, S., Huang, Z.,
  Karpathy, A., Khosla, A., Bernstein, M., Berg, A.C., Fei-Fei, L.: Imagenet
  large scale visual recognition challenge. International Journal of Computer
  Vision  \textbf{115}(3),  211--252 (Dec 2015).
  \doi{10.1007/s11263-015-0816-y},
  \url{https://doi.org/10.1007/s11263-015-0816-y}

\bibitem{schou2016algorithms}
Schou, J., Farkas, J.: Algorithms, interfaces, and the circulation of
  information: Interrogating the epistemological challenges of facebook. KOME:
  An International Journal of Pure Communication Inquiry  \textbf{4}(1),
  36--49 (2016)

\bibitem{STUMPF2009639}
Stumpf, S., Rajaram, V., Li, L., Wong, W.K., Burnett, M., Dietterich, T.,
  Sullivan, E., Herlocker, J.: Interacting meaningfully with machine learning
  systems: Three experiments. International Journal of Human-Computer Studies
  \textbf{67}(8),  639 -- 662 (2009).
  \doi{https://doi.org/10.1016/j.ijhcs.2009.03.004},
  \url{http://www.sciencedirect.com/science/article/pii/S1071581909000457}

\bibitem{tullio2007works}
Tullio, J., Dey, A.K., Chalecki, J., Fogarty, J.: How it works: a field study
  of non-technical users interacting with an intelligent system. In:
  Proceedings of the SIGCHI Conference on Human Factors in Computing Systems.
  pp. 31--40. ACM (2007)

\bibitem{Woodruff:2018:QEP:3173574.3174230}
Woodruff, A., Fox, S.E., Rousso-Schindler, S., Warshaw, J.: A qualitative
  exploration of perceptions of algorithmic fairness. In: Proceedings of the
  2018 CHI Conference on Human Factors in Computing Systems. pp. 656:1--656:14.
  CHI '18, ACM, New York, NY, USA (2018). \doi{10.1145/3173574.3174230},
  \url{http://doi.acm.org/10.1145/3173574.3174230}

\bibitem{ozgobek}
Özgöbek, O., Shabib, N., Gulla, J.: Data sets and news recommendation. In:
  Workshops Proceedings of the 24th ACM Conference on User Modeling,
  Adaptation, and Personalization. vol.~1181, pp. 5--12 (01 2014)

\end{thebibliography}
\end{document}